\documentclass[aps,twocolumn,superscriptaddress,showpacs]{revtex4}
\usepackage{bm}
\usepackage{graphicx}

\begin{document}
\bibliographystyle{revtex}
\draft


\title{Magnetic field effect on the dielectric constant of glasses : an  evidence of disorder within tunneling barriers ? }

\author{J. Le Cochec}
\author{F. Ladieu}
\email[]{ladieu@drecam.cea.fr}
\affiliation{DSM/DRECAM/LPS, C.E.Saclay, 91191 Gif sur Yvette Cedex, France}
\author{P. Pari}
\affiliation{DSM/DRECAM/SPEC, C.E.Saclay, 91191 Gif sur Yvette Cedex, France }

\date{\today}

\begin{abstract}
The magnetic field dependence of the low frequency dielectric 
constant $\epsilon_{r}(H)$ of a structural glass $a-\mathrm{SiO_{2+x}C_{y}H_{z}}$ was studied from $400  $ $\mathrm{mK}$ 
to $50  $ $\mathrm{mK}$ and for $H$ up to $3  $ $\mathrm{T}$. Measurement of 
 \textit{both} the real and the imaginary parts of $\epsilon_{r}$ is used to eliminate
 the difficult question of keeping constant the temperature of the sample while increasing $H$: a non-zero $\epsilon_{r}(H)$ dependence is reported in the same range as that 
one very recently reported on multicomponent 
glasses. \textit{In addition } 
to the recently proposed explanation based on interactions, the 
reported $\epsilon_{r}(H)$ is interpreted quantitatively as 
a consequence of \textit{the disorder} lying within the nanometric barriers 
of the elementary tunneling systems of the glass.

\end{abstract}

\maketitle

\section{Introduction}

Since the seventies \cite{Zeller}, it is well established that at low temperature $T$, the
properties of amorphous solids differ strongly from their 
crystaline counterparts. This is currently explained \cite{Anderson} within the 
tunneling two-level system (TLS) model, stating that the only relevant degrees 
of freedom at low $T$ are particles going back and forth between 
two neighboring sites of equilibrium. Indeed, the TLS model accounts for various properties of glasses below $\sim 1  $ $\mathrm{K}$. A closer look, however, reveals that the TLS model 
often does not quantitatively account for the measured quantities. Since the late eighties, considerable efforts 
have been made to include interactions between TLS's in order to 
explain such discrepancies \cite{Granaan} and justify the TLS model 
from first principles \cite{Klein}, \cite{Burin} or, at least, to give a 
more consistent picture of TLS's \cite{Wurger}, \cite{Kettemann}. However,
 the lack of TLS microscopic information allowed to consider interactions 
either to be weak \cite{Burin} or strong \cite{Wurger}, 
\cite{Kettemann}, \cite{Strehlow98} : these opposite starting points echo
 Leggett 
and Yu's question : "Is there anything really tunneling in 
glasses ?" \cite{Leggett}. 

The purpose of this work is to gain some microscopic information on  
TLS's by studying the 
magnetic field effect on a glassy property, namely 
the $1  $ $\mathrm{kHz}$ dielectric constant $\epsilon_{r}$. Indeed, for a charge $Q$,
 the tunneling transparency of a barrier
 of size $a$ is strongly
 modified by magnetic fields $H$ of the order 
of $H_{c} = {\hbar \over{Q a^2}}$ : 
this is due to the ($H$ dependent) difference of quantum phase picked up by the
 various paths under the barrier \cite{Kettemann}.
However, setting the usual TLS parameters, $a \simeq 0.2  $ $\mathrm{nm}$ and $Q=e$,
 with $e$ the electronic charge, leads to 
 $H_{c} \simeq 10^{4}  $ $\mathrm{T}$, a value so large that it 
seemed to explain Frossati \textit{et al}'s results \cite{Frossati} reporting 
no detectable $H$ effects on $\epsilon_{r}$. Recently, however, 
detectable $\epsilon_{r}(H)$ effects were reported below $80  $ $\mathrm{mK}$ (and 
down to $1  $ $\mathrm{mK}$) on a multicomponent glass 
\cite{Strehlow98}, \cite{Strehlow00}.
 Very strikingly,  $\epsilon_{r}(H)$ is peaked around $H_{c} \simeq 0.03  $ $\mathrm{T}$,
 a very low value accounted for by using interactions : 
at low $T$, it was argued that interactions correlate  a large number  
$N \gg 1$ of TLS's
 and that the resulting spectrum resembles that of a single 
TLS with a charge $Q=Ne$ (and a renormalized tunneling energy $\Delta_{0}$). 
In this work, a detectable $\epsilon_{r}(H)$ effect is reported on a $a-\mathrm{SiO_{2+x}C_{y}H_{z}}$  
glass for $50  $ $\mathrm{mK}\le T\le 400  $ $\mathrm{mK}$ and $H$ up to $3  $ $\mathrm{T}$. Even if our glass is not really a model glass (such as the various forms of vitreous silica, e.g. Suprasil), the    
 $\epsilon_{r}(H)$ dependence reported here shows that these effects are not restricted to the multicomponent glass studied previously \cite{Strehlow98}, \cite{Strehlow00}, favoring the idea that such effects are general in structural glasses. Furthermore our work shows, as stated previously \cite{Ahn}, that such measurements  demand special experimental care since $\epsilon_{r}(H)$ effects are quite small: $\delta \epsilon_{r}'(H) /\epsilon_{r}'$ lies typically in the $10^{-6} -10^{-5}$ range ($\epsilon_{r}'$ is the real part of $\epsilon_{r}$). This is why, after a brief description of the experimental setup, it is explained how to overcome the difficult question of keeping $T$ constant with great accuracy,  while increasing $H$. The resulting 
$\epsilon'_{r}(H)$ effect is then accounted for by using numerical calculations of the $H$ effect
 on the tunneling transparency of a \textit{disordered} nanometric 
barrier. As explained below, this additional mechanism should not contradict
 the one based upon interactions.

\section{Experimental setup}

The sample was deposited on a $a-\mathrm{SiO_{2}}$ substrate as follows: 
i) a $0.1   $ $\mathrm{\mu m}$ thick Cu layer was first evaporated, 
ii) a $L=0.8  $ $\mathrm{\mu m}$ thick $a-\mathrm{SiO_{2+x}C_{y}H_{z}}$ layer was then deposited by a 
$13  $ $\mathrm{MHz}$ plasma vapor technique using TetraEthylOrthoSilane mixed with 
He 
\cite{Boutard} 
iii) a $0.08  $ $\mathrm{\mu m}$ thick Cu layer, followed by a $0.04  $ $\mathrm{\mu m}$ Au  
layer was evaporated. Both electrodes are $3  $ $\mathrm{mm}$ large $10  $ $\mathrm{mm}$ long ribbons 
recovering each other along $\simeq 3  $ $\mathrm{mm}$ (i.e., the capacitance  
surface is $S \simeq 9  $ $\mathrm{mm}^2$) :
 the opposite ends of the electrodes are free of glass deposit allowing an  
ohmic contact to be made with a Cu wire glued 
with Ag paste. These two Cu wires are soldered to the inner wires 
of cryogenic coaxial cables going from room temperature to the cold 
copper box embedding the sample : this box is related to the mixing chamber 
of the dilution fridge by a thermal impedance and contains a 
$\mathrm{RuO_{2}}$ resistive thermometer whose resistance will be called $\theta_{\mathrm{RuO_2}}$ throughout this work (to avoid any confusion with the parallel resistance $R$ of the glass sample). The sample substrate was glued with varnish on a 
copper sample holder for thermalization. To decrease possible 
$H$-dependent heating in the electrodes, $H$ was set parallel to the
 electrodes (i.e., perpendicular to the $1  $ $\mathrm{kHz}$ measuring field
$\cal E$). Semi-rigid home-made coaxial cables were installed at the top of 
the fridge and directly plugged in the Andeen-2500A capacitance bridge. The capacitance bridge was used in its parallel option, yielding $C\propto \epsilon_{r}'$ and $G \propto \epsilon_{r}''$, but throughout this work it was chosen to report on $C$ and $R=1/G$ due to the great similarity of the $C(T)$ and $R(T)$ curves (see Fig. 1) : this strongly confirms that the reported $R$ at low $T$ is not some parasitic edge effect but has the same origin that the reported $C$, namely the TLS's.

From $300  $ $\mathrm{K}$ to $4.2  $ $\mathrm{K}$, the capacitance $C$ typically halves,
 while the parallel resistance $R$ increases by a factor
 $\simeq 100$. Below $4.2  $ $\mathrm{K}$, $C$ decreases, reaches its 
minimum for $T_{rev.}$ and then increases below $T_{rev.}$ before reaching a 
ultra low $T$ saturation value $C_{sat}$. According to the standard TLS 
model the $C$ decrease above $T_{rev.}$ is due to progressive freezing of 
the diagonal (or relaxational) part of the response, while the $C$ 
increase below $T_{rev.}$ comes 
from the induced off-diagonal (or resonant) part of the susceptibility : 
this effect enlarges as $T$ decreases as do all quantum effects.

\section{Obtaining $\delta \epsilon'_{r}(H)$}
\subsection{Proof that a non-zero  $\delta \epsilon'_{r}(H)$ exists}

\begin{figure}
\includegraphics[height=7cm, width=8.5cm]{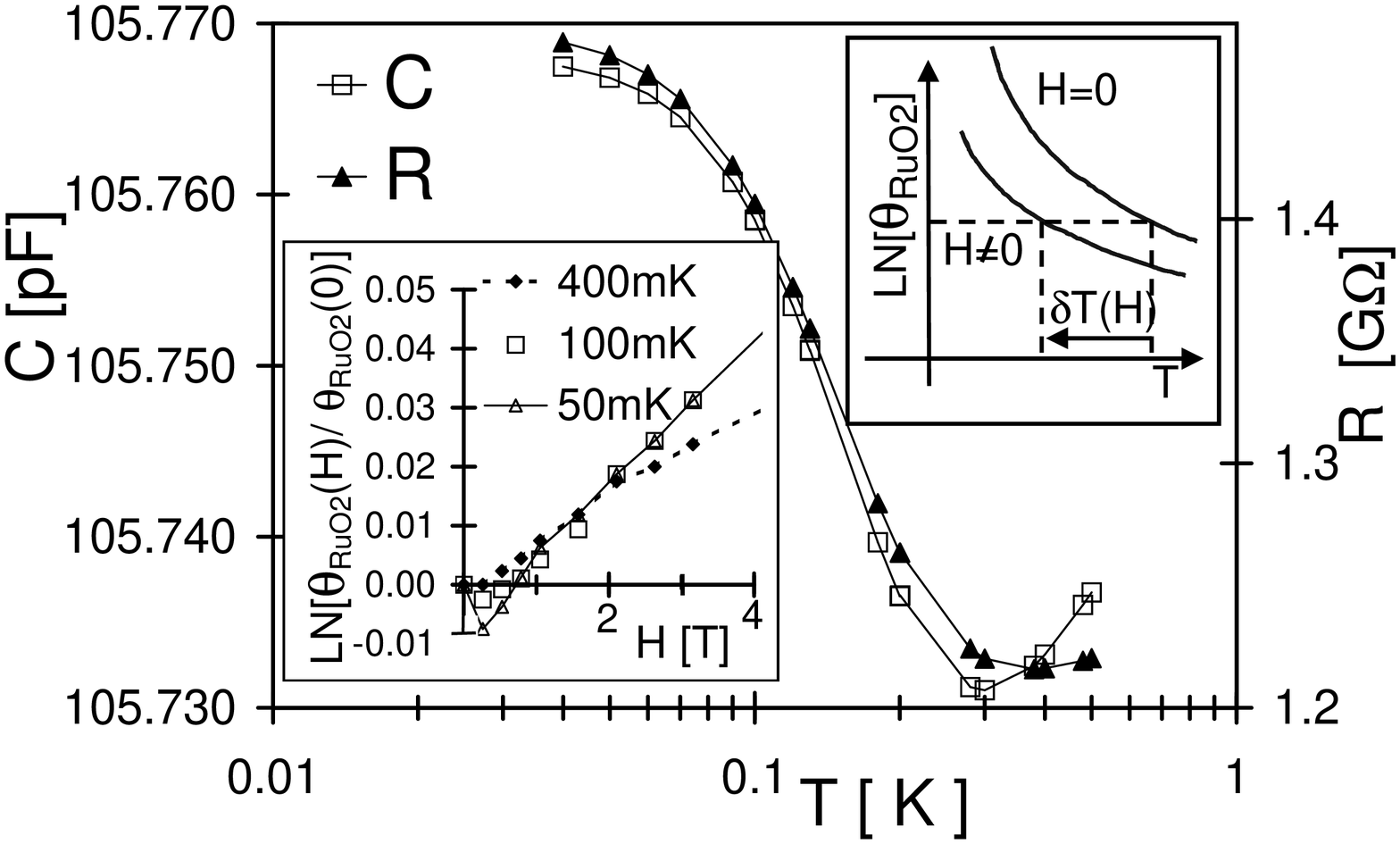}
\caption{$C(T)$ (left axis) and $R(T)$ (right axis) when a $1  $ $\mathrm{kHz}$ measuring field ${\cal E}=3.75  $ $\mathrm{MV/m}$ is applied to the glass sample. \textit{ Lower inset}: 
$\theta_{\mathrm{RuO_2}}(H)$ dependence measured assuming that the ${\cal P}(T)$ relationship does not depend on $H$ (see text). \textit{ Upper inset}: Schematic effect of a negative $\mathrm{RuO_2}$ magnetoresistance : increasing $H$ while keeping $\theta_{\mathrm{RuO_2}}$ constant leads to a small decrease of $T$, i.e.,  $\delta T (H)<0$.}
\label{Fig. 1}
\end{figure}

Figure 1 shows $C(T)$ and $R(T)$ curves when a $1  $ $\mathrm{kHz}$ field ${\cal E}= 3V/0.8  $ $\mathrm{\mu m} =3.75   $ $\mathrm{MV/m}$ is used. The high value of 
$T_{rev.} \simeq 0.3  $ $\mathrm{K} \simeq 8T_{rev.}({\cal E }\to 0)$ reveals
 that such fields correspond to the strongly 
nonlinear regime \cite{Osheroff}. Such large fields were used 
with a large integration time ($100  $ $\mathrm{s}$) so as to reduce 
the bridge uncertainties $c_{noise}$ (resp. $r_{noise}$) over $C$ 
(resp. $R$) measurements down to $c_{noise}/C = \pm 10^{-7}$
 (resp.  $r_{noise}/R = \pm 10^{-5}$). Large fields were also used 
in \cite{Strehlow00}. 

The second key point to investigate small $\epsilon_{r}(H)$ effects is 
to be able to increase $H$ while keeping $T$ accurately constant.
 It demands either to set up a $H$-independent thermometry, which is extremely
 difficult over an extended $T$ range (it may have been
 the way used in \cite{Strehlow98}, \cite{Strehlow00}); or,
 at least, to correct $T$ of the magnetoresistance of the thermometer. 
However, such a correction cannot be done with great accuracy since the standard 
technique for measuring the $\theta_{\mathrm{RuO_2}}(H)$ dependence is as follows  \cite{Sanquer} : 
\textit{i)} For $H=0$ the relationship between the 
power $\cal P$ injected in the fridge and the resulting $T$ is first 
recorded ; \textit{ii)} For a given $\cal P$, 
$H$ is slowly raised to a given value, and the measured  
$\theta_{\mathrm{RuO_2}}(H)-\theta_{\mathrm{RuO_2}}(0)$ at thermal equilibrium is attributed to the 
magnetoresistance of the 
thermometer : the corresponding results are shown in the lower inset of Fig. 1 and lead to corrections of the order of $2-5\%$ on $T$ when $H$ is raised up to a few $\mathrm{T}$. 
However, these  $\theta_{\mathrm{RuO_2}}(H)$ measurements lie on the key assumption that the ${\cal P}(T)$ relationship does not depend at all on $H$ : this cannot hold accurately, e.g., due to the dissipation of energy resulting 
from the vibrations of the fridge within the magnetic field. These small and complicated effects  
cannot be ignored when seeking 
small $\epsilon_{r}(H)$ effects which, translated into a thermal 
equivalent, amount to an effective variation of $T$ lying in the $0.1-5\%$ range (see {\bf B)}\textit{2)} below).

\begin{figure}
\includegraphics[height=7cm, width=8.5cm]{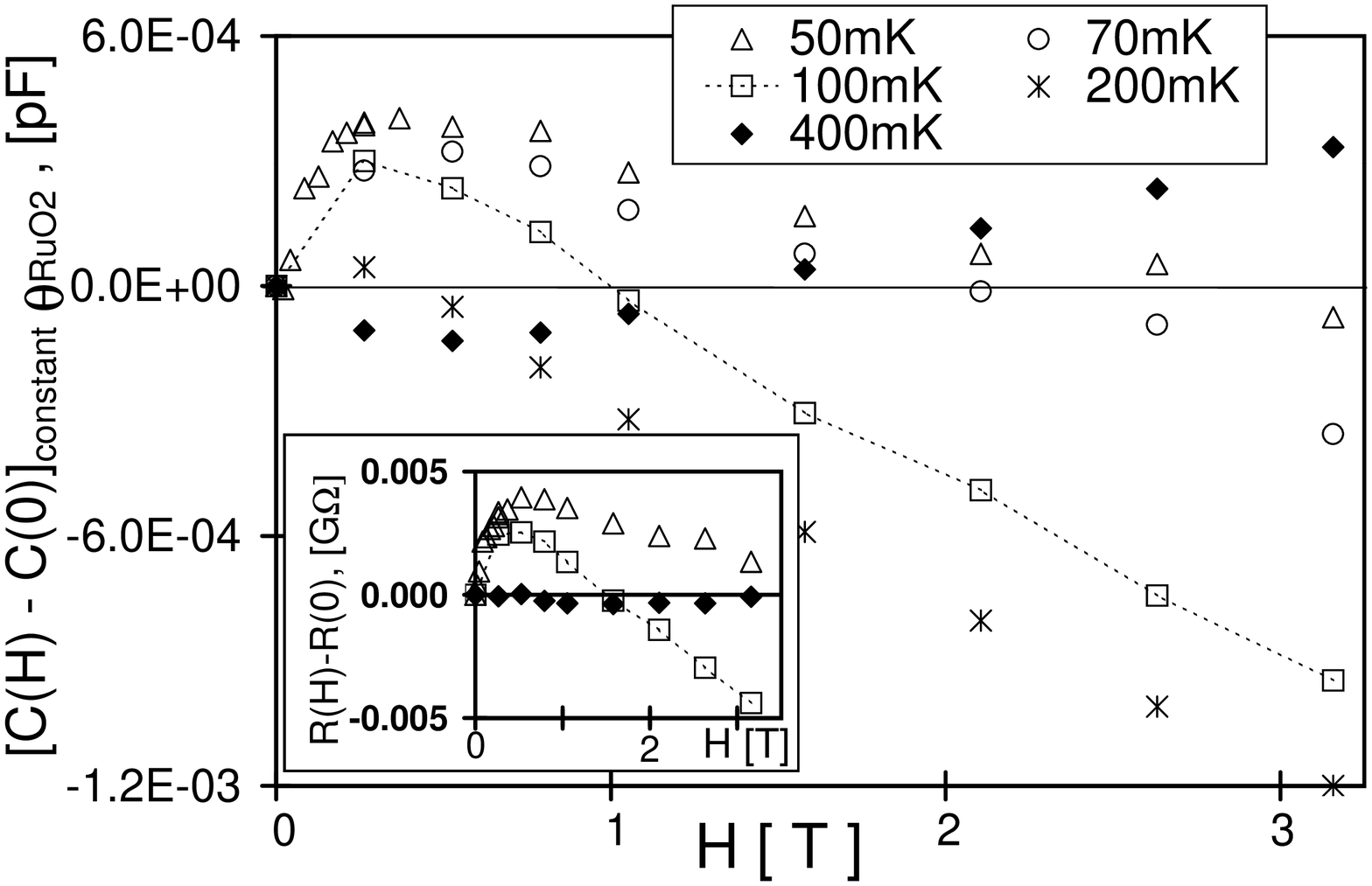}
\caption{ $\delta C(H)$ measured by keeping $\theta_{\mathrm{RuO_2}}$ constant while increasing $H$ (the various $T$ labeling the curves are at $H=0$). \textit{ Inset}: Corresponding data for $\delta R(H)$ (same symbols). The fact that $\delta R(H)$ and $\delta C(H)$ do not vanish at the same $H$ cannot be explained by using \textit{ only} $\delta T(H)$ 
effects but implies that a $\delta \epsilon_{r} (H)$ dependence exists in the sample. The error bars $c_{noise}=\pm 10^{-5}  $ $\mathrm{pF}$ and $r_{noise}= \pm 10^{-5}  $ $\mathrm{G\Omega}$ are much smaller than the data symbols.}
\label{Fig. 2}
\end{figure}

Going back to the $C,R$ measurements, the fact that both ${\cal P}(T)$ and $\theta_{\mathrm{RuO_2}}(T)$ relationships sligthly depend on $H$ implies that $T$ cannot be made  
accurately constant when $H$ is increased. For practical reasons, the 
 $C(H),R(H)$ measurements were made letting $\cal P$ 
automatically adjust so as to keep the $\theta_{\mathrm{RuO_2}}$ value fixed at
 any $H$ value : raising $H$ while keeping $\theta_{\mathrm{RuO_2}}$ constant leads to a small shift 
$\delta T(H)$ of the temperature of the experiment as
 depicted in the upper inset of Fig. 1.

The measured variations $\delta C(H) = [C(H)-C(0)]_{constant\  \theta_{\mathrm{RuO_2}}}$ are displayed in Fig. 2 for various $T$ 
values measured at $H=0$. The key point is that the $\delta T(H)$ dependence 
cannot explain all the qualitative features of Fig. 2. For example, whatever $\delta T(H)$
 may be, if  $\delta \epsilon_{r}(H)$ were zero, $\delta C(H)$ and $\delta R(H)$ would vanish  
at the same $H$ : this is clearly contradicted by the $50  $ $\mathrm{mK}$, $100  $ $\mathrm{mK}$ and $400  $ $\mathrm{mK}$ data reported on Fig. 2. \textit{This proves that a $\epsilon_{r}(H)$ 
dependence exists in our sample} but that an accurate correction of the 
$\delta T(H)$ effects must be done on the data of Fig. 2. This correction cannot be done with the data of the inset of Fig. 1 since, as above evoked, the sought $\epsilon_{r}(H)$ is smaller than the accuracy of the $\delta T(H)$ estimate.

\subsection{Obtaining the order of magnitude of $\delta \epsilon'_{r}(H)$}

At this step, the existence of a detectable $\delta \epsilon'_{r}(H)$ dependence 
is established but an accurate suppression of $\delta T(H)$
 effects is still missing. It is shown here that using \textit{ both}
 $\delta R(H)$ and $\delta C(H)$ allows to draw from Fig. 2 the order of magnitude of 
$\delta \epsilon'_{r}(H)$, i.e., to know its value within a scale factor 
$0.1 < \eta < 10$ (see Eq. (2) below) : this interval might seem to be very large, but we will see in the next section that the physical interpretation only involves $\ln(\eta)$.

\subsubsection{Overview of the method}

The key idea is that since the relative variations of all the involved quantities are small, first order expansions are accurate.
 Defining $\delta X(H) = X(H)-X(0)$ for any quantity $X$ measured as a function of $H$ keeping the $\theta_{\mathrm{RuO_2}}$ value constant, one gets :

$$\delta C(H) = \left( \frac{\partial C}{\partial T}\right)\delta T(H) + \delta C_{int.}(H) ,  \eqno(1a)$$

where the measured $\delta C(H)$ is shown on Fig. 2, 
$\delta C_{int.}(H)$ is proportionnal to the sought intrinsic variation of 
$\delta \epsilon_{r}'(H)$ and throughout this work, all the partial derivatives with respect to $T$ are taken at $H=0$.
 Introducing  
the corresponding quantity $\delta R_{int.}(H)$, 
 an equation similar to Eq. (1a) is obtained for $R$ measurements. 

$$\delta R(H) = \left( \frac{\partial R}{\partial T}\right)\delta T(H) + \delta R_{int.}(H) . \eqno(1b)$$

Combining Eqs. (1a) and (1b) 
 allows to \textit{accurately} eliminate $\delta T(H)$, and 
to get :

$$\delta C_{int.}(H) = \eta \left( 
\frac{
\left(\frac{\partial C}{\partial T}\right)
 } {
\left(\frac{\partial R}{\partial T}\right)
 }
\delta R(H) - \delta C(H) \right) , \eqno(2a)$$

 with $\eta = \frac{1}{(\alpha -1 )}$ and $\alpha$ is defined by :

$$\alpha  = \frac{\delta R_{int.}(H)}{\delta C_{int.}(H)}\frac{\left(\frac{\partial C}{\partial T}\right) }{\left(\frac{\partial R}{\partial T}\right)} . \eqno(2b)$$

 $\alpha$ is introduced here 
as an attempt to eliminate one of the three unknown quantities among $\delta T(H)$,
 $\delta C_{int.}(H)$ and $\delta R_{int.}(H)$ since only two quantities, 
 $\delta C(H)$ and $\delta R(H)$, are measured. 
  $\alpha$ cannot be predicted since, for 
strong $\cal E$, a comprehensive treatment of the TLS dynamics, including quantum 
coherence effects, is still missing : at present, even the $C(T)$ curve is not well
 accounted for \cite{Osheroff}. However, within the framework of the standard TLS model,
 $\alpha$ and $\eta$ \textit{ cannot depend on} $H$ since, as we show below, 
$\Delta_{0,max}$ is the only parameter of the TLS model which might slightly 
depend on $H$ : neglecting, as above, any second order terms, both the real and the 
imaginary parts of $\delta \epsilon_{r}(H)$ (i.e., both $\delta C_{int.}(H)$ and $\delta R_{int.}(H)$) are expected 
to be proportionnal to the \textit{same} small 
$\delta \Delta_{0,max}(H)/ \Delta_{0,max}$ (see Eq. (4)).

There is no similar argument stating that, generally, $\alpha$ and $\eta=1/(\alpha-1)$ cannot
 depend on $T$. It is shown in the section A of the Appendix, 
by using the shape of the curves of Fig. 2 and quite general arguments,
 that $\eta$ lies, \textit{ at any reported} $T$, 
 in $[0.1,10]$, an interval which will turn out to be 
"small" enough to neglect any $\eta(T)$ variations 
(see section {\bf IV}). 
Basically, it is shown in the Appendix that the $\delta T(H)$ influence on 
the measured $\delta C(H)$ and $\delta R(H)$ enlarges as $H$ is increased : the quasilinear decrease of $\delta C(H), \delta R(H)$ seen in Fig. 2 at large $H$ thus comes from $\delta T(H)$ effects (at $400  $ $\mathrm{mK}$ it turns to a quasilinear increase due to the fact that $\partial C/ \partial T >0$ contrarily to the case of lower $T$ data).
  
\subsubsection{The most salient features of $\delta C_{int.}(H)$}

 Since $0.1 < \eta < 10$, the order of magnitude of $\delta C_{int.}(H)$ is obtained by setting  $\eta=1$ and disregarding any possible $\eta(T)$ variations within this interval : the resulting 
transformation of the data of Fig. 2 through 
Eq. (2a) is shown on Fig. 3. $\delta C_{int.}(H)$ is positive for the 
lowest temperatures (except for three low fields at $50  $ $\mathrm{mK}$, see below) and 
negative at higher $T$, i.e., in the vicinity of $T_{rev.}$. 

The $200  $ $\mathrm{mK}$ data were not reported on Fig. 3 since they lie very close to zero. 
Indeed, it seemed more important to report, in the inset of Fig. 3, 
on $\delta C(H)$ measured just 
at $T_{rev.}$ (measured with a larger field ${\cal E} = 8.75  $ $\mathrm{MV/m}$ to further increase the signal to noise ratio). For the inset of Fig. 3, at low $H$, the $\delta T(H)$ contribution (see Eq. (1a)) to  $\delta C(H)$ is negligible since  
 $\partial C/ \partial T=0$ just at $T_{rev.}$ (at large $H$ this is no longer true due to the increase of $\delta T(H)$, yielding the quasilinear increase of $\delta C(H)$). The important point here is the decrease of $\delta C(H,T_{rev.})$ at
 low $H$ : it further confirms that $\delta C_{int.}(H)$ is negative in the vicinity of $T_{rev.}$ and that $\eta>0$ at $400  $ $\mathrm{mK}$. 
Note that the results of 
Fig. 3 were reproduced on a second similar sample.

\begin{figure}
\includegraphics[height=7cm, width=8.5cm]{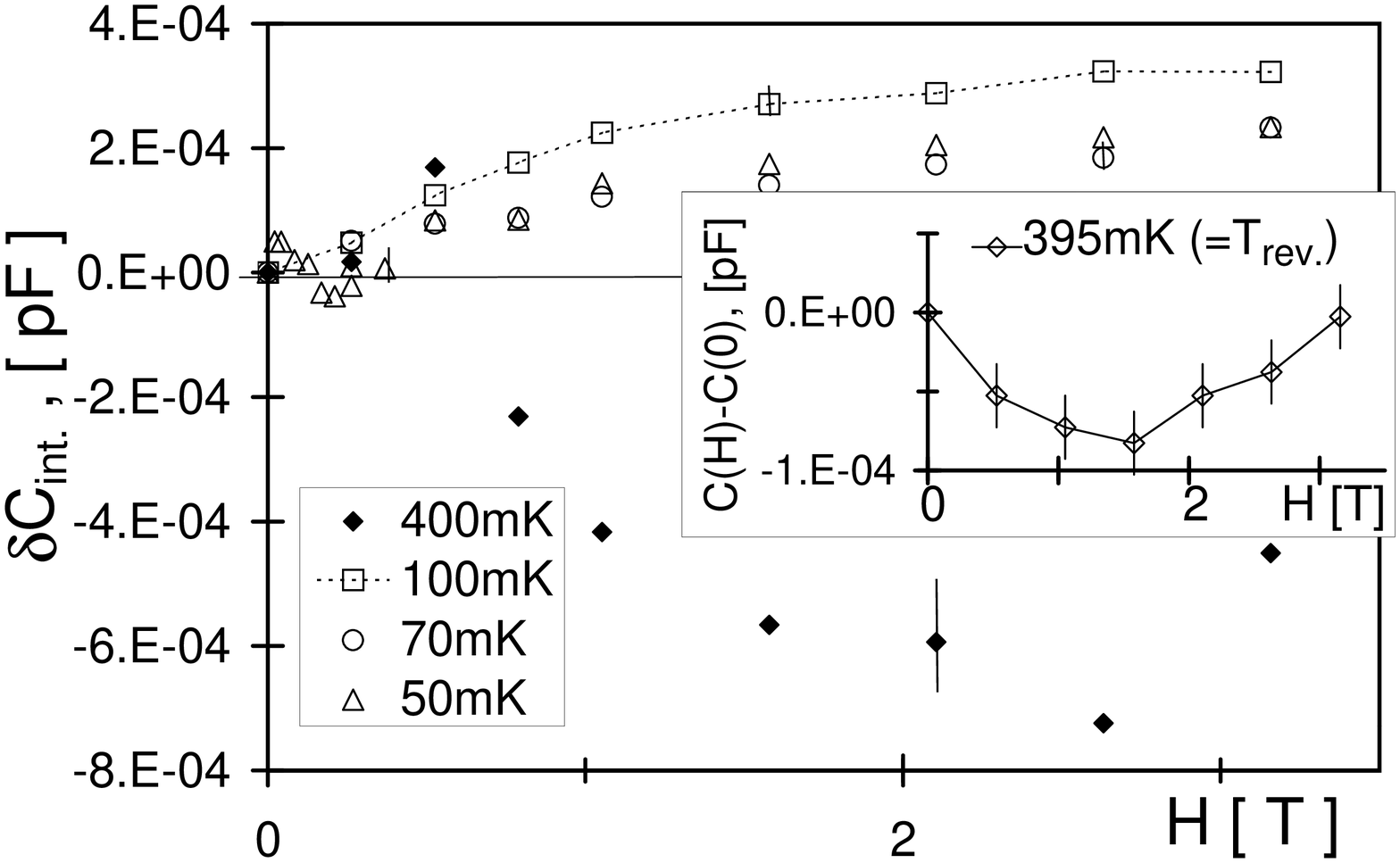}
\caption{ $\delta C_{int.}(H) \propto \delta \epsilon'_{r}$ drawn from Fig. 2 data by using Eq. (2a) with $\eta=1$. For clarity the error bar (see Eq. (3)) is reported only for one $H$ at each $T$ : it is much stronger at $400  $ $\mathrm{mK}$ than below $100  $ $\mathrm{mK}$ where error bars hardly exceed the size of the data symbols. \textit{ Inset}: $\delta C(H)$ measured just at $T_{rev.}$ (with a larger field ${\cal E} = 8.75  $ $\mathrm{MV/m}$ to further increase the signal to noise ratio) : due to the $\partial C/ \partial T =0$ at $T_{rev.}$ the use of Eq. (2a) is not necessary at low $H$ (at higher $H$, $\delta T(H)$ is no longer negligible and produces the $\delta C(H)$ increase). Comparison of
 the data at $T_{rev.}$ to those at $400  $ $\mathrm{mK}$ supports the idea that $\eta>0$ at $400  $ $\mathrm{mK}$.}
\label{Fig. 3}
\end{figure}

Let us translate the  data of Fig. 3 to a thermal equivalent by using  $\delta T_{int.C}= \delta C_{int.}(H)/({\partial C \over{\partial T}})$ (remember that $\delta C_{int.}(H)$ are calculated by setting $\eta=1$) : it is found that 
$\delta T_{int.C}(H)$ reaches
 at most $7\%$ of $T$ at $50  $ $\mathrm{mK}$, $2\%$ at $70$ $\mathrm{mK}$, $1\%$ at $100  $ $\mathrm{mK}$ and $4\%$ at $400  $ $\mathrm{mK}$. \textit{ Beyond chemical differencies
 between glasses, this might explain why, with 
an uncertainty on $T$ of $\pm 5\%$, previous studies reported undetectable $\epsilon_{r}(H)$ effects} \cite{Frossati}. Adding 
quadratically the uncertainties on both terms of the right hand-side of 
Eq. (2a), 
 the final uncertainty $c_{err}$ about the 
$\delta C_{int.}(H)$ reported on Fig. 3 reads : 

$$c_{err} = \eta \sqrt{ (c_{noise})^{2} + (r_{noise}{\partial C \over{\partial T} } )^{2}
({\partial R \over {\partial T} })^{-2} } . \eqno(3)$$

 On Fig. 3, $c_{err}$ was reported for each $T$ :  due to the strong decrease of $\partial R/ \partial T$, it is much stronger at 
$0.4  $ $\mathrm{K}$ than at low $T$ where it hardly exceeds the symbol size.
 However, even at $0.4  $ $\mathrm{K}$, $c_{err}$ amounts 
to an uncertainty over $T$ of $\pm 0.5\%$, ten times smaller than 
in \cite{Frossati}. 

Let us note that $\delta C_{int.}(H)/c_{err}$ does not depend on the chosen 
$\eta =1$ : the trends of $\delta C_{int.}(H)$ reported on Fig. 3 are
 thus reliable at each $T$, even if the possible $\eta(T)$ dependence forbids any accurate comparison of $\delta C_{int.}(H)$ at different $T$. Comparing Fig. 3 with Strehlow's 
results \cite{Strehlow00}, our $\delta \epsilon'_{r}(H)$ is
 somewhat smaller but lies in the same range. 
As in \cite{Strehlow00}, our $\delta \epsilon_{r}'(H)$ \textit{ might}
 be peaked around $H \simeq 0.02  $ $\mathrm{T}$ for our lowest $T$, but this effect is too close to 
our $c_{err}$ to be systematically studied. 

\section{Physical interpretation}

Let us move to the interpretation of Fig. 3. First, note that, due to the 
$T$-dependence of the data of Fig. 2 and Fig. 3, 
the reported $\delta \epsilon'_{r}(H)$ 
can be attributed neither to the coaxial cables nor to the "matrix" surrounding 
TLS's. Furthermore, the change of sign of $\delta C_{int.}(H)$ in the vicinity
 of $T_{rev.}$ excludes a role of the electrode-glass interfaces and 
further confirms that TLS's are responsible for the 
reported $\delta \epsilon'_{r}(H)$. Last, it is shown in the section B of the Appendix  that spin effects, if any, can be ruled out at low enough $T$, i.e., well below $T_{rev.}$ : this is why we will focus on the positive $\delta\epsilon'_{r}(H)$ observed at $T\le 200$ $\mathrm{mK}$. 

\subsection{Going from $\delta C_{int.}(H)$ to $\delta \Delta_{0,max}(H)$}

For the part of the capacitance $C_{TLS}$ coming from TLS's, the usual result \cite{Osheroff} below $T_{rev.}$ can thus be used : 
$C_{TLS} = {\cal C}\ln{(E_{max}/k_{B}T)}$ where $k_{B}$ is 
Boltzmann constant, $E_{max} = \sqrt{\Delta_{max}^{2} + \Delta_{0,max}^{2}}$ is 
the maximum interlevel spacing of TLS's, and the constant
 ${\cal C} \simeq 0.03  $ $\mathrm{pF}$ is determined by fitting the
 increase of $C(T)$ of Fig. 1 
 below $T_{rev.}$ (details are given below). 

Since both $\Delta$ (see section B of the Appendix) and $\cal C$ \cite{note3} are $H$-independent,
\textit{ the reported} $\delta \epsilon'_{r}(H)$ 
\textit{is interpreted as a}
 $\Delta_{0, max}(H)$ \textit{dependence}, yielding :

$${2\delta C_{int.}(H) \over {\cal C}}  = 
{{\delta \Delta_{0,max}(H)} \over{\Delta_{0,max}(0)}}
=\ln{( {{\Delta_{0,max}(H)}\over {\Delta_{0,max}(0)}} )} , \eqno(4)$$
where the standard assumption $\Delta_{max}=\Delta_{0,max}$ is responsible 
for the factor $2$ in the left hand-side of Eq. (4).

 Let us note that the expression of $C_{TLS}$ used to derive Eq. (4) only holds accurately between 
$90  $ $\mathrm{mK}$ and $200  $ $\mathrm{mK}$ since, at lower $T$, the $C(T)$ curve tends to saturate. This failure of the fit at low $T$ is due to the fact that the above expression of $C_{TLS}$ is well established only in the linear regime, and that a comprehensive treatment for the strong $\cal E$ used here is still missing. However, this formula should capture the essential part of the physics since it expresses that, well below $T_{rev.}$, the TLS susceptibility is due to the quantum transitions coherently induced by $\cal E$. Since, physically, it is likely that tunneling plays a key role in the nonlinear $C_{TLS}$ at low $T$, Eq. (4) should finally hold, up to a multiplicative prefactor which can be disregarded just like $\eta$ (see below).

\begin{figure}
\includegraphics[height=7cm, width=8.5cm]{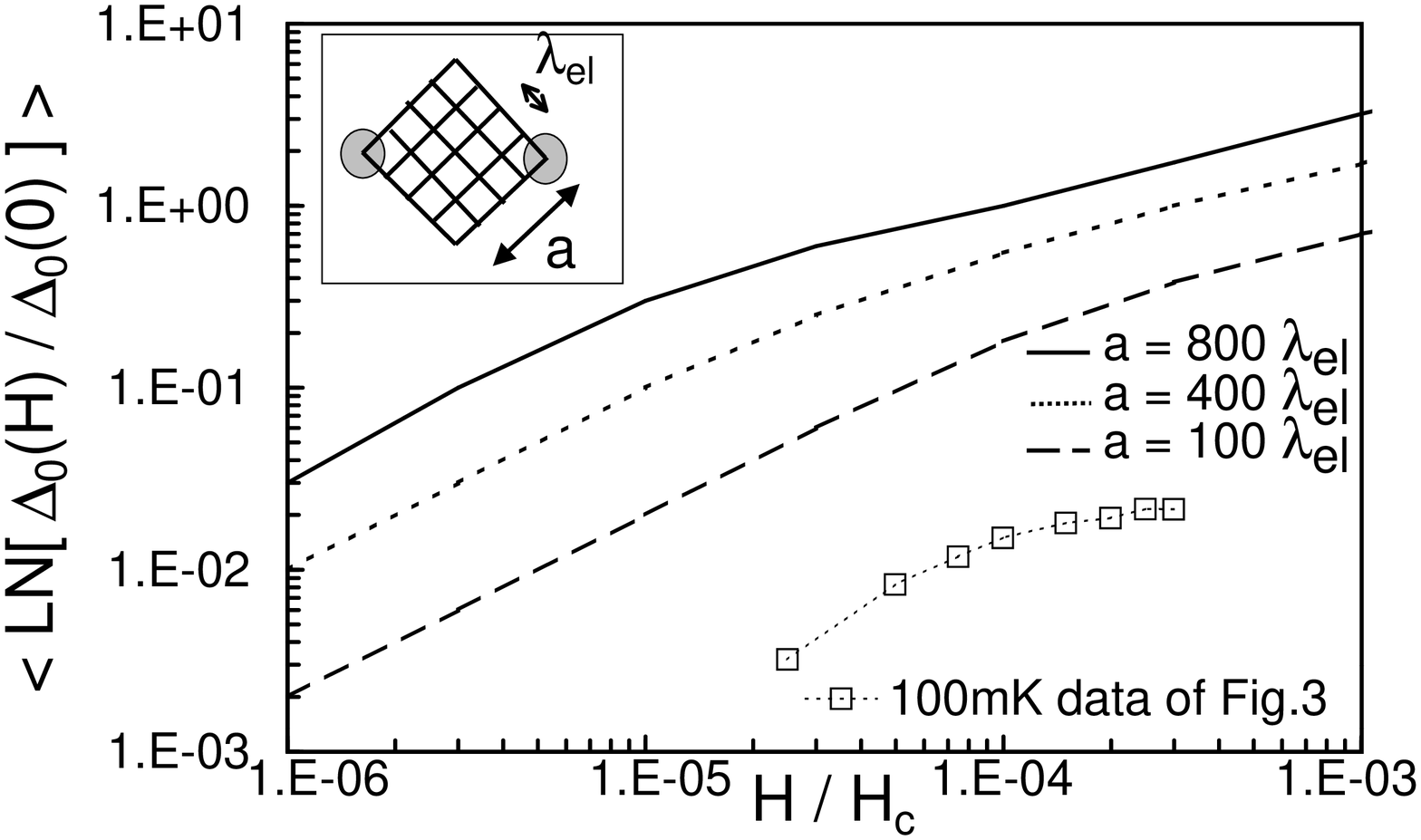}
\caption{ \textit{ Inset}: Schematic view of the disorder lying within
 the tunnel barrier of size $\sim a$ between the two sites 
(gray circles) of a TLS : the potential fluctuations are
 modeled by a set of "impurities" of interspacing $\lambda_{el}$ whose 
energies are drawn at random either well above or well below that of 
the TLS. \textit{ Main figure}: $\Delta_{0}$ results from the coherent sum 
of all the paths linking the TLS sites : due to the additional quantum 
phase introduced by $H$ for each path, $<\ln \Delta_{0}>$ increases with $H$,
 the $H^2$ dependence at very low $H$ 
(unreported) turning continuously to a $H^{0.5}$ behavior at higher
 fields (ensemble averaging over the possible "impurities" configurations 
is indicated by $<\ >$ and $Hc= { {\hbar} \over {e \lambda_{el}^{2}}}$). The
 three curves for $a/\lambda_{el} \ge 100$ are calculations drawn from Medina
 \textit{et al}'s work \cite{Medina} and the $100  $ $\mathrm{mK}$ data of Fig. 3 are reported by 
using Eq. (4). Both the sign and the sublinear behavior of our data are 
well accounted for by the calculations. Extrapolation from the logarithmic 
behavior at large $a/\lambda_{el}$ gives $a/\lambda_{el} \simeq 5-10$ for our
 sample, yielding an elementary dipole $p_{0} \sim 10  $ $\mathrm{D}$ not that far from
 the one drawn from echo measurements.}
\label{Fig. 4}
\end{figure}

\subsection{A possible microscopic explanation of $\delta \Delta_{0,max}(H)$}

A $\Delta_{0,max}(H)$ dependence means that the tunnel transparency is affected 
by $H$. This is well known in another kind of disordered 
insulators, namely Anderson insulators (e.g. lightly doped semiconductors) 
where electronic states are localized due to potential disorder. At 
very low frequency,
 transport in Anderson insulators 
occurs by hopping on a fractal percolating path strongly 
affected by $H$.
 $H$ effects are smaller at high frequencies where the conducting path is 
broken since only the fastest hops contribute  
: note that, since they are disconnected from each other, each of these fast back-and-forth 
hops is a TLS. 
In both limits, $H$ effects were accounted for \cite{Sanquer}, \cite{Levy} 
by studying the tunnel transparency $\Delta_{0}$ 
between two sites \textit{separated by a disordered barrier} 
depicted in the inset of Fig. 4 : tunneling between the two 
sites of the (quasisymmetric) 
TLS separated by the distance $a$ is strongly affected by
 the potential disorder coming from the structural disorder within the 
tunneling barrier itself. This is modeled by a three dimensional network of 
"impurities" (whose unit length is the elastic mean free path 
$\lambda_{el}$) and $\Delta_{0}$ results from the coherent sum of all 
quantum paths along the impurity network : at $H=0$, a path picks up a 
phase $-1$ each time it diffuses on an impurity whose energy $e_{i}$ is smaller than that of the TLS. Since the $e_{i}$ are drawn at random, 
$\Delta_{0}$ must be first averaged over disorder 
(noted hereafter with $<\ >$). 

Using this model to account for the measured $\delta \epsilon'_{r}(H)$ is obviously an oversimplification. Indeed, TLS's might be more complicated than
 a \textit {unique} charge going back and forth between two sites separated by $a$: a  TLS could be a group of charges, a 'molecule', and more generally any charged species with two spatially separated potential minima. However, we show now that this model accounts for most of the salient features of the low-$T$ $\delta \epsilon'_{r}(H)$ reported in Fig. 3. 
 
\subsection{Accounting for the data by using the disordered tunneling barrier model}

Figure 4 is drawn from Medina \textit{et al}'s calculations 
\cite{Medina} and shows 
that $<\ln{\Delta_{0}}>$ \textit{increases} with $H$
 (in Fig. 4 $H_{c}= {\hbar \over{e \lambda_{el}^2}}$), 
due to the additional quantum phase introduced by $H$. 
Using Eq. (4) with ${\cal C} =0.03  $ $\mathrm{pF}$ and $\lambda_{el}=0.2  $ $\mathrm{nm}$, the positive  
 $\delta \epsilon'_{r}(H)$ reported at $100  $ $\mathrm{mK}$ on Fig. 3 can be displayed on Fig. 4. Comparison with the calculations, made for large $a/\lambda_{el.}$, reveals that both \textit{the sign and the sublinear dependence}  
$\delta \epsilon'_{r}(H) \sim H^{0.6 \pm 0.2}$ observed on the data
 are compatible with disorder effects. Moreover, using the roughly logarithmic decrease 
of $<\delta \ln (\Delta_{0}(H))>$ with $a/\lambda_{el}$, 
the experimental data are quantitatively compatible with $a \simeq 5 \lambda_{el} \simeq 1  $ $\mathrm{nm}$, 
yielding an elementary dipole $p_{0} \simeq 10   $ $\mathrm{D}$ \cite{note2}: this is the good order of magnitude with respect to the $4  $ $\mathrm{D}$ 
value drawn from pulse echo experiments \cite{echo}. Last,   
let us note that, once reported on Fig. 4, the $50$ $\mathrm{mK}$ and $70$  $\mathrm{mK}$ data lie very close to the $100$ $\mathrm{mK}$ data, yielding the same $\delta \epsilon'_{r}(H) \sim H^{0.6 \pm 0.2}$ as well as the same order of magnitude for $a/\lambda_{el.}$ : these low $T$ data were not reported on Fig.4 since the expression of $C_{TLS}$ used to derive eq. (4) does not hold precisely below $90$ $\mathrm{mK}$.

To summarize, the microscopic model presented here allows to fit quantitatively the data with very reasonnable parameters. This is due to the extreme $H$-sensitivity of tunneling through a \textit{disordered} barrier which yields non negligible effects even for $H \ll H_{c}$. Indeed, suppressing disorder and taking $a = \lambda_{el}=0.2$ $\mathrm{nm}$, would lead, by standard tunneling
 calculations, to 
$<\delta \ln (\Delta_{0}(H))> \simeq -(H/H_{c})^{2} \simeq -10^{-8}$, 
i.e., to an \textit{undetectable quadratic negative} effect, at odds with experiments.
 
Despite its success in accounting for our data, a limit of our mechanism is that it yields a monotonous $H$-behavior of $\delta \epsilon'_{r}(H)$, i.e., it  cannot account for the peaked structure around $H=0.03  $ $\mathrm{T}$ reported in \cite{Strehlow00}. However, even if our scenario is based on disorder effects,
 it might not contradict the interaction picture used 
in \cite{Strehlow00}. Indeed, the disorder explanation 
lies upon a $\Delta_{0,max}(H)$ effect with a very large 
$H_{c}\simeq 10^{4}  $ $\mathrm{T}$ while the interaction mechanism involves a $\Delta_{0,min}(H)$ effect with a very small  
$H_{c}\simeq 10^{-2}  $ $\mathrm{T}$ (due to the very large number $N$ of correlated TLS's).
 In the Fig. 3 of \cite{Ahn},
 it seems that collective tunneling can be viewed simply as a simultaneous 
tunneling of $N$ elementary dipoles. If it is so, both mechanisms should 
simply add up due to their very different $H_{c}$ . It is well admitted \cite{Strehlow00} 
that averaging problems in the interaction picture demand further progress.
 It should explain, e.g. why, in \cite{Strehlow00}, $\delta \epsilon'_{r}(H)$ is peaked
 around a $H_{c}$ field \textit{ decreasing} when $T$ increases
, which \textit{naively} amounts to a counter-intuitive
 \textit{increase} of $N$ when $T$ increases.

\section{Conclusion}

To summarize, a $\epsilon_{r}(H)$ dependence was 
shown in a $a-\mathrm{SiO_{2+x}C_{y}H_{z}}$ glass for $H\sim 1  $ $\mathrm{T}$ and $T\leq 0.4  $ $\mathrm{K}$.
 This demanded to decrease the  
$T$ uncertainty to less than $0.5\%$ : beyond  
chemical differences between glasses, this might explain why previous 
studies reported undetectable $\epsilon_{r}(H)$ effects. At low $T$,
the reported  $\epsilon_{r}(H)$ effects were interpreted 
assuming that tunneling is affected by some disorder
 within the elementary tunneling barriers of size $a\simeq 1  $ $\mathrm{nm}$. This scenario might simply add up to the interacting one previously proposed.

\acknowledgements{Many thanks to P. Ailloud (CNRS/LPS) and P. Trouslard (CEA/INSTN/LVdG)
for samples realisation, and to P. Forget for cryogenic help. 
Useful discussions with D. Boutard, M. Ocio, M. Rotter, J. Joffrin and J.-Y. Prieur
 are greatly acknowledged.}


\section{Appendix}

\subsection{Proof that $\eta$ lies in the interval $[0.1;10]$}

It is argued here that, at any reported $T$, $\eta$ lies in the interval 
$[0.1,10]$. Defining the exponents $s$ and $t$ by $\delta T(1  $ $\mathrm{T}\le H \le 3  $ $\mathrm{T}) \sim H^t$ and $\delta C_{int.}(1  $ $\mathrm{T}\le H \le 3  $ $\mathrm{T}) \sim H^s$, we will first assume that $t-s \ge 0.3$ and show, from the data of Fig. 2, that this implies $0.1 \le \eta \le 10$ at any reported $T$. Reciprocally, assuming  $0.1 \le \eta \le 10$, we will show, by using the trends of Fig. 4 and Fig. 1 that $t-s > 0.3$. Note that $s$ and $t$ are not necessarily integers, due to subtle disorder averaging effects, such as those  studied in Medina \textit{et al}'s work \cite{Medina} (see section {\bf IV}). 

Since we must precisely compare the relative importances in eqs. (1a)-(1b) of the first term containing $\delta T(H)$ and the second one containing $\delta \epsilon_{r}(H)$, it is more convenient to translate the measured $\delta C(H)$ and the sought $\delta C_{int.}(H)$ into thermal equivalents by introducing $\delta T_{C}(H) = \delta C(H)/({\partial C\over \partial T})$  
and $\delta T_{int.C}(H) = \delta C_{int.}(H)/({\partial C\over \partial T})$. Eq. (1a) thus amounts to: 

$$\delta T_{C}(H) = \delta T(H) + \delta T_{int.C}(H) , \eqno(A1)$$

while defining 
the corresponding quantities $\delta T_{R}(H)$ and $\delta T_{int.R}(H)$ 
 for $R$ measurements yields from Eq. (1b): 

$$\delta T_{R}(H) = \delta T(H) + \delta T_{int.R}(H) . \eqno(A2)$$

With these definitions note that the definition of $\alpha$ given in Eq. (2b) becomes 
$\alpha = \delta T_{int.R}/\delta T_{int.C}$ with still $\eta=1/(\alpha-1)$. 

\subsubsection{Assuming $t-s>0.3$ implies $0.1\le \eta \le 10$ at any reported $T$}

\textit{i)} \textit{ $\eta <10$ (i.e., $\alpha>1.1$) }. Consider the magnitudes of the 
$\delta C(H)$ and $\delta R(H)$ bumps reported on Fig. 2 : for 
example at $50  $ $\mathrm{mK}$, these positive bumps around $H=0.5  $ $\mathrm{T}$ amount to 
$\delta T_{C} \simeq -6  $ $\mathrm{mK}$ 
and to $\delta T_{R} \simeq -7.6  $ $\mathrm{mK}$. These two values are ten times larger than that  
 derived from the small negative magnetoresistance of 
the $\mathrm{RuO_2}$ thermometer (see inset of Fig. 1) and, even 
 if the inset data cannot be accurately trusted, these $\delta T_{C},\delta T_{R}$
  values are anyway too large to be entirely attributed to the $\mathrm{RuO_{2}}$ thermometer \cite{note}. 
This means that both $\delta T_{int.C}(H)$ and 
$\delta T_{int.R}(H)$ are negative at $50  $ $\mathrm{mK}$, i.e., $\alpha >0$ at $50  $ $\mathrm{mK}$ and  
both $\delta C_{int.}(H)$ and 
 $\delta R_{int.}(H)$ positive in the Tesla range : with the above definition of $s$, $\delta C_{int.}(H)$ and $\delta R_{int.}(H)$ are expected to be proportionnal 
to the same $H^{s}$ with positive coefficients. Thus $\delta C_{int.}(H)$ and 
$\delta R_{int.}(H)$ \textit{increase} with $H$ : the (quasilinear) decrease seen on 
Fig. 2 at large $H$
 thus comes from $\mathrm{RuO_2}$ effects, i.e., $\delta T(H) \propto H^{t}$ and $t>s$.
 Setting these $H$ expansions in Eqs. (A1),(A2) leads to $\alpha = (H_{R}/H_{C})^{t-s}$
 with $H_{R}$ the magnetic field where $\delta R(H)$ vanishes and $H_{C}$ 
the corresponding one for $\delta C(H)$ : at $50  $ $\mathrm{mK}$, $H_{R} \simeq 4.2  $ $\mathrm{T}$ and $H_{C} \simeq 2.9  $ $\mathrm{T}$, which with the above assumption $t-s>0.3$ leads to $\alpha >1.1$, i.e., $\eta < 10$.
 Finally, the same argument can be made at $T>50  $ $\mathrm{mK}$ where $H_{R}/H_{C}$ ratios are larger,
 reaching $2.5$ at $400  $ $\mathrm{mK}$. 

\textit{ii)} \textit{ $\eta >0.1$ (i.e., $\alpha<10$). }
 The assumption $t>s+0.3$ implies that the low field region $H \lesssim 1  $ $\mathrm{T}$ is that where the  relative importance of 
$\delta T_{int.C}$ (resp. $\delta T_{int.R}$) in the measured 
 $\delta T_{C}$ (resp. $\delta T_{R}$) is the largest. In this range where 
$\delta C(H)$ and $\delta R(H)$ have the same sign, the 
data of Fig. 2 correspond, for any given $T$, to values of $\delta T_{C}$ and $\delta T_{R}$ which differ by less
 than a factor $2$. This means that $\delta T_{int.C}$ and
 $\delta T_{int.R}$ are of the same order of magnitude 
(this might be seen also in \cite{Strehlow00}), hence $\alpha < 10$ i.e, $\eta > 0.1$: this result 
holds even at large $H$ since $\alpha$ and $\eta$ do not depend on $H$.

\subsubsection{Assuming $0.1\le \eta \le 10$ implies $t-s>0.3$ }

This assumption on $\eta$ allows Fig. 3 and Fig. 4 to be drawn from Fig. 2, yielding $s=0.6\pm0.2$ for $H\gtrsim 1  $ $\mathrm{T}$. Besides, the data of the lower inset of Fig. 1 about $\theta_{\mathrm{RuO_2}}(H)$ effects
 amount to $t \gtrsim 1.2$ for $1  $ $\mathrm{T} \lesssim H \le 3  $ $\mathrm{T}$ (note that we do not rely here on the precise values of $\theta_{\mathrm{RuO_2}}(H)$ effects but only on their general trend). Hence, the data ensure $t-s>0.3$.

\subsection{Ruling out spin effects at low $T$}

Due to the lack 
of microscopic information on TLS's, spin effects cannot be ruled out at all $T$.
Indeed, if each TLS contains a single electronic spin, the non negligible 
Zeeman energy $\delta E_{Z}(H) \simeq 0.7  $ $\mathrm{K/T}$ comes into play : each TLS becomes a four-level system. This  
complicates greatly the $\epsilon'_{r}$ calculation in the range $T \gtrsim T_{rev.}$   
where the relaxational (i.e., "thermodynamic") contribution plays an important role. This is why the 
negative $\delta \epsilon'_{r}(H)$ close to $400$ $\mathrm{mK}$   
has not been interpreted and the efforts were  
focused on the positive $\delta \epsilon'_{r}(H)$ at low $T$. Indeed,
 well below $T_{rev.}$, the TLS susceptibility is driven by the
 transitions \textit{coherently }induced by $\cal E$, and such quantum 
transitions are forbidden between states of  different spins : TLS's can thus be separated into two \textit{independent} 
subclasses of a given spin state whose susceptibilities can be added.
 Thus, for a given spin state, the asymmetry energy $\Delta$ between the two 
potential wells of given TLS 
is effectively independent of $\delta E_{Z}(H)$. 

%
%


\begin{thebibliography}{99}
\bibitem{note} If the magnetoresistance of the $\mathrm{RuO_2}$ thermometers were so large to lead 
to a $-6.2  $ $\mathrm{mK}$ shift of the temperature at $50  $ $\mathrm{mK}$, this would have dramatic consequences 
in the vast majority of experiments where 
the $H$ dependence of physical quantities are measured
 while neglecting the magnetoresistance of the $\mathrm{RuO_2}$ : the latter should not lead to corrections over $T$ greater than $5\%$, contrarily to the $12.4\%$ value corresponding to 
$6.2  $ $\mathrm{mK}$ for $50  $ $\mathrm{mK}$.
\bibitem{note3} In the linear regime, $\cal C$ is proportionnal to two 
$H$-independent parameters of the TLS model : $p_0$ the elementary TLS dipole, and the density of states.
\bibitem{note2} Since $\delta C_{int.}
\propto \eta$, the more general result is $a/\lambda_{el} \simeq 5 + \ln(\eta)$, yielding negligible corrections whatever $\eta$ in its 
$[0.1,10]$ interval.
\end{thebibliography}
\end{document}